\documentclass[a4paper,fleqn,usenatbib]{mnras}

\usepackage{mathptmx}

\usepackage[T1]{fontenc} 
\usepackage{ae,aecompl}

\usepackage{graphicx}	
\usepackage{amsmath}	
\usepackage{amssymb}	

\usepackage{gensymb}

\title[Bowen emission from Aquila X-1]{Bowen emission from Aquila X-1: evidence for multiple components and constraint on the accretion disc vertical structure}

\author[F. Jim\'{e}nez-Ibarra et al.]{F. Jim\'{e}nez-Ibarra,$^{1,2}$\thanks{E-mail: felipeji@iac.es}
T. Mu\~{n}oz-Darias,$^{1,2}$
L. Wang,$^{3}$
J. Casares,$^{1,2,5}$\newauthor
D. Mata S\'{a}nchez,$^{1,2}$
D. Steeghs,$^{3}$
M. Armas Padilla$^{1,2}$
and P. A. Charles$^{4,5}$
\\
$^{1}$Instituto de Astrof\'{i}sica de Canarias, V\'{i}a L\'{a}ctea, La Laguna, E-38205, Santa Cruz de Tenerife, Spain\\
$^{2}$Departamento de Astrof\'{i}sica, Universidad de La Laguna, E-38206, Santa Cruz de Tenerife, Spain\\
$^{3}$Department of Physics, University of Warwick, Gibbet Hill Road, Coventry CV4 7AL, UK\\
$^{4}$Department of Physics and Astronomy, University of Southampton, Southampton SO17 1BJ, UK\\
$^{5}$Department of Physics, Astrophysics, University of Oxford, Denys Wilkinson Building, Keble Road, Oxford OX1 3RH, UK}

\date{Accepted XXX. Received YYY; in original form ZZZ}

\pubyear{2017}

\begin{document}

\label{firstpage}
\pagerange{\pageref{firstpage}--\pageref{lastpage}}
\maketitle

\begin{abstract}

We present a detailed spectroscopic study of the optical counterpart of the neutron star X-ray transient Aquila X-1 during its 2011, 2013 and 2016 outbursts. We use 65 intermediate resolution GTC-10.4m spectra with the aim of detecting irradiation-induced Bowen blend emission from the donor star. While Gaussian fitting does not yield conclusive results, our full phase coverage allows us to exploit Doppler mapping techniques to independently constrain the donor star radial velocity. By using the component N \textsc{iii} 4640.64/4641.84  \AA\ we measure $K_{\mathrm{em}}$ = 102 $\pm$ 6 km $\mathrm{s^{-1}}$. This highly significant detection ($\gtrsim 13\sigma$) is fully compatible with the true companion star radial velocity obtained from near-infrared spectroscopy during quiescence. Combining these two velocities we determine, for the first time, the accretion disc opening angle and its associated error from direct spectroscopic measurements and detailed modelling, obtaining $\alpha = 15.5$ $ ^{+ 2.5}_{-5}$ deg. This value is consistent with theoretical work if significant X-ray irradiation is taken into account and is important in the light of recent observations of GX339-4, where discrepant results were obtained between the donor's intrinsic radial velocity and the Bowen-inferred value. We also discuss the limitations of the Bowen technique when complete phase coverage is not available.


\end{abstract}

\begin{keywords}
X-rays: binaries -- accretion, accretion discs -- stars: neutron 
\end{keywords}



\section{Introduction}

Low-mass X-ray binaries (LMXBs) are stellar systems comprising a low-mass star ($\lesssim 1$ M$_{\sun}$) which transfers matter on to a compact object (either a neutron star or a black hole) via an accretion disc. Depending on their long-term behaviour we can distinguish between persistent systems, those constantly accreting material at high rates, and transient systems, which spend the majority of their lives in a dormant, quiescent state. The latter show sporadic outbursts lasting weeks to years, when their X-ray luminosity increases typically to $\geq$ 10 per cent of the Eddington luminosity, becoming at least as bright as persistent sources (e.g. \citealt{2006AdSpR..38.2675V}; \citealt{2011BASI...39..409B}; \citealt{2016LNP...905...65F} for reviews).

Classical dynamical studies of LMXBs rely on the spectroscopic detection of the companion star at optical and/or near-infrared wavelengths (e.g. \citealt{2003astro.ph..8020C}; \citealt{2014SSRv..183..223C}). However, the optical counterparts of most transient LMXBs are faint in quiescence ($V > 22$--$23$) and to obtain phase-resolved spectra during this epoch is not always an easy task. In addition, even if optical spectroscopy is feasible, the companion star features might be veiled by the disc emission  \citep[e.g.][]{2015MNRAS.454.2199M}. The situation is even more complex during outburst, when the radiation coming from the outer accretion disc generally dominates even in the near-infrared \citep{2015MNRAS.449L...1M}.

The so-called Bowen technique was developed to measure the orbital motion of the donor star using narrow fluorescence lines within the Bowen blend region, a typical spectral feature of LMXBs in outburst. In a few key systems, it has been shown that these narrow components originate as a result of the X-ray reprocessing on the irradiated side of the donor. Thus, a lower limit to the radial velocity of the companion ($K_{2}$) can be established by measuring Doppler shifts from these emission components \citep[i.e $K_{\mathrm{em}}$,][see also \citealt{2007MNRAS.379.1637M} for Bowen blend time-lags consistent with the donor star]{2002ApJ...568..273S}. Narrow emission lines in the Bowen region have been observed in a dozen or so objects \citep{2008AIPC.1010..148C,2009PASP..121..935M}, which led to the first estimate of the system parameters in three transient systems in outburst \citep[e.g.][]{2017MNRAS.466.2261W} and several persistent sources \citep[e.g.][]{2006MNRAS.373.1235C}.

\defcitealias{2017MNRAS.464L..41M}{MS2017} 

Aquila X-1 (Aql~X-1) is a neutron star transient LMXB that has shown recurrent outbursts since its discovery by \cite{1967Sci...156..374F}. It has a short outburst recurrence time of $\sim$ two years, being one of the most prolific and widely studied X-ray transients (e.g. \citealt{2010ApJ...716L.109M}; \citealt{2014MNRAS.438.2634C}; \citealt{2014MNRAS.443.3270M}). During the 2004 outburst, \cite{2007MNRAS.375.1463C} constrained $K_{2}$ by using the Bowen technique, and obtained $K_{\mathrm{em}}=247 \pm 8 $ km s$^{-1}$. This result is no longer compatible with the dynamical solution derived from near-infrared spectroscopy taken during quiescence \citet[][hereafter \citetalias{2017MNRAS.464L..41M}]{2017MNRAS.464L..41M}. The discrepancy could result from spurious emission coming from localised regions of the disc that contribute to the Bowen blend \citep{2004MNRAS.348..100H}. Similarly, a recent spectroscopic study of the black hole transient  GX~339-4 \citep{2017arXiv170804667H} during one of its short quiescence epochs ruled out a companion star origin for the $K_{\mathrm{em}}$ reported by \cite{2003ApJ...583L..95H}. Both the Aql X-1 and GX~339-4 outburst studies have in common a limited phase coverage.

\defcitealias{2005ApJ...635..502M}{MD2005} 

In this paper we present phase-resolved optical spectroscopy of Aql~X-1 taken during the 2011, 2013 and 2016 outbursts. We take advantage of a complete phase coverage to exploit the Doppler mapping technique in order to accurately determine $K_{\mathrm{em}}$, deriving a value which is now fully compatible with the near-infrared measurement. Subsequently, we use $K_{2}$ from \citetalias{2017MNRAS.464L..41M} and the ratio between  $K_{\mathrm{em}}$ and $K_{2}$ ($K_{\mathrm{c}}$; as defined in \citealt[][hereafter \citetalias{2005ApJ...635..502M}]{2005ApJ...635..502M}) to determine the opening angle of the accretion disc, therefore constraining its vertical extent during outburst.

\section{Observations and data reduction}\label{sec:obs&data}

We obtained phase-resolved spectroscopy of Aql X-1 with the Optical System for Imaging and low-Intermediate Resolution Integrated Spectroscopy (OSIRIS) located in the Nasmyth-B focus of the 10.4m Gran Telescopio Canarias (GTC) at the Observatorio del  Roque de los Muchachos (La Palma, Spain). We used the optical grism R2500V (0.8 \AA\ pix$^{-1}$)  centred at 5185 \AA\ with  spectral coverage between  4500--6000 \AA\ and two different slit widths (0.8 and 1 arcsec), which yielded a velocity resolution of  $132.3 \pm 0.2$  and $172.9 \pm 0.2$ km s$^{-1}$, respectively [full width at half-maximun (FWHM)]  at 4624.27 \AA.

A total of 65 spectra were acquired during three different outburst episodes, in 2011, 2013 and 2016, as detailed in Table \ref{journal}. Note that in one case we have averaged four spectra of 300s in two bins (600 s) to increase the signal-to-noise ratio, but the effective integration time is still short compared to the binary orbital period ($<$0.02 in phase). The spectra were taken over 19 different epochs [see Fig.~\ref{fig:compareoutburst} (bottom) ] with seeing in the range 0.75--1.45 arcsec.

\setlength{\tabcolsep}{3pts}

\begin{table}
	\centering
	\caption{Journal of observations GTC/OSIRIS spectroscopy of Aql X-1}
	\label{journal}
	\begin{tabular}{ccccc} 
		\hline
		Date & UT interval & No. & Exp.  time & Orb. phase interval* \\
		\hline
		26/10/2011 & 20:53--21:25 &4 & 600s & 0.58--0.61\\
         29/11/2011 & 19:50--20:01& 2 & 600s & 0.58--0.60\\
		
		25/06/2013 & 03:18--03:54 & 4 & 671s & 0.74-0.77\\
		05/07/2013 & 03:28--04:03 & 4& 671s & 0.42--0.45\\
		09/07/2013 & 01:24--01:59 & 4 & 671s& 0.38--0.41\\
		11/07/2013 & 01:44--03:08 & 8 & 671s & 0.93--0.00\\
		14/07/2013 & 02:07--02:43 & 4 & 671s & 0.75--0.78\\

		03/08/2016 & 01:04--02:04 & 6 & 700s & 0.23--0.28\\
		
		05/08/2016 & 00:47--02:00  & 8 & 600s & 0.75--0.81 \\
		05/08/2016 & 23:47--23:57 & 2 & 600s & 0.96--0.97\\

		07/08/2016 & 00:04--00:15 & 2 & 600s & 0.24--0.25 \\
		07/08/2016 & 23:39--23:54 & 2 & 600s & 0.48--0.50\\

		11/08/2016 & 23:03--23:14 & 2 & 600s & 0.52--0.53\\
		13/08/2016 & 22:59--23:10 & 2 & 600s & 0.05--0.06\\
		18/08/2016 & 21:28--21:44 & 2 & 600s (2 $\times$ 300s) & 0.30--0.32\\
		21/08/2016 & 22:34--22:44 & 2 & 600s & 0.16--0.17\\
		24/08/2016 & 22:24--22:34 & 2 & 600s & 0.95--0.96\\
		27/08/2016 & 21:52--22:03 & 2 & 600s & 0.72--0.73\\
		29/08/2016 & 21:19--21:43 & 3 & 685s & 0.23--0.25\\
		\hline 
	\end{tabular}
\begin{flushleft}
*for orbital ephemeris see text section \ref{gaussian_fit}.
\end{flushleft}
\end{table}

The data were de-biased and flat-fielded using \textsc{iraf}\footnote{\textsc{iraf} is distributed by National Optical Astronomy Observatories, operated by the Association of Universities for Research in Astronomy, Inc., under contract with the National Science Foundation.}  standard routines. The pixel-to-wavelength calibration was made using a regular HgAr+Ne+Xe arc lamp exposure taken on each observing block. Before extracting the spectra, cosmic rays were removed from the 2-D spectra using \textsc{L.A.Cosmic} \citep{2001PASP..113.1420V}. Velocity drifts due to instrumental flexure effects were corrected ($<$40 km s$^{-1}$) using the \textsc{molly} software and the OI 5577.338 \AA\ line from the sky spectra.

\section{Results}\label{sec:results}

\begin{figure}
\begin{center}
 \begin{minipage}[b]{\columnwidth}
    \includegraphics[width=\columnwidth]{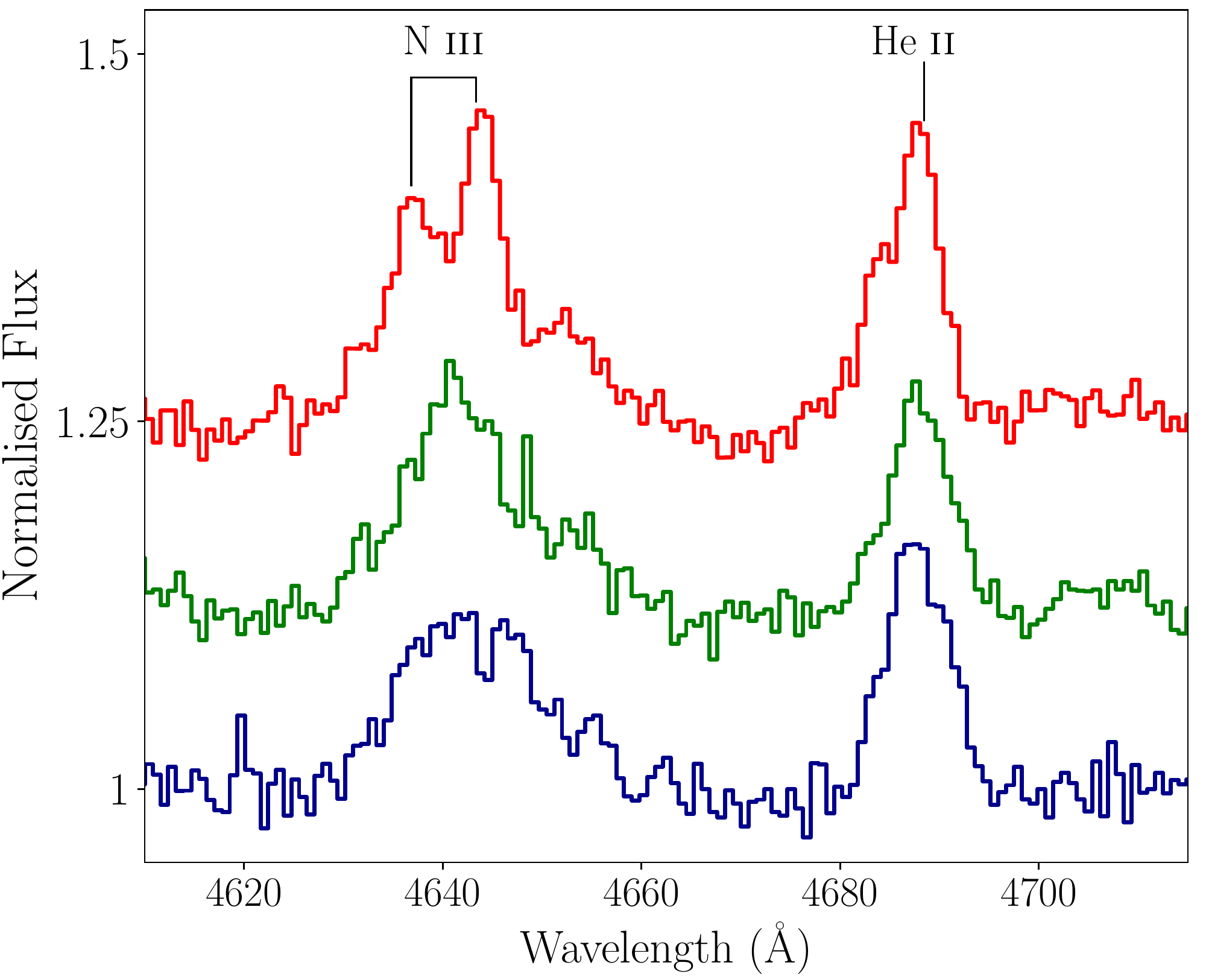}
 	\includegraphics[width=\columnwidth]{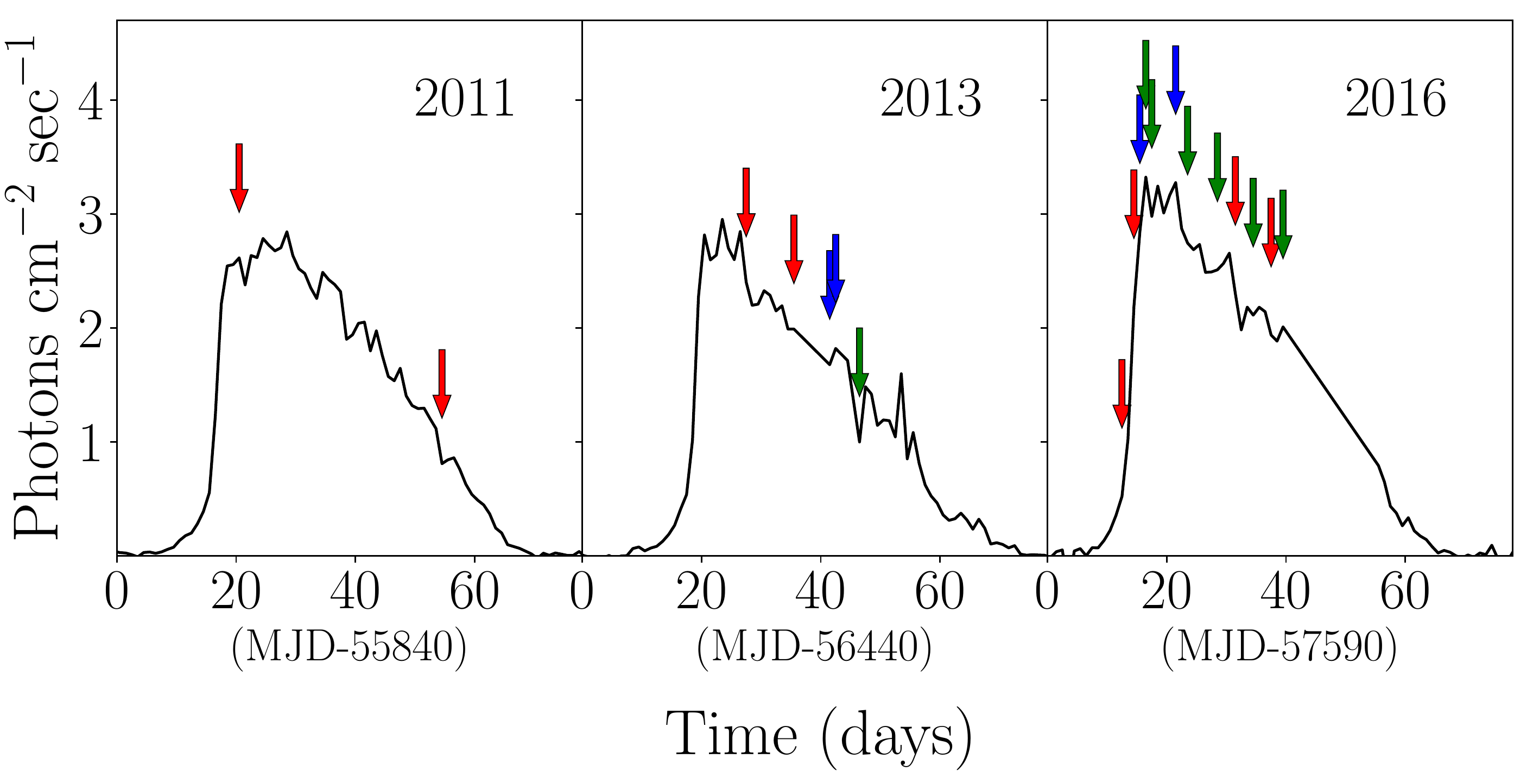}
  \end{minipage}
\end{center}
 
 \caption{Top: examples of normalised  spectra of Aql X-1 in the Bowen blend region for each of the three groups defined in the text (section \ref{sec:results}): spectra with securely identified N \textsc{iii} narrow emission lines (red line; positive offset applied), spectra showing unidentified narrow components (green line; positive offset applied) and spectra with no narrow components within the Bowen blend (blue line). Bottom: X-ray light curves of Aql X-1 during the 2011, 2013 and 2016 outbursts obtained with the Monitor of All-sky X-ray Image (\textit{MAXI}). Arrows indicate the time of our GTC observing blocks using the same colour code as that of the top panel.}
    \label{fig:compareoutburst}
\end{figure}

The Bowen blend complex is detected in all 65 GTC spectra. However, its relative strength and shape show strong variability. We can distinguish 3 groups: (i) 38 spectra show narrow lines that we could unambiguously identify as N \textsc{iii} 4634.13 and  N \textsc{iii} 4640.64/4641.84  \AA; (ii) 19 spectra show narrow features, but do not enable a clear identification of the N \textsc{iii} components, and  (iii) 8 spectra do not show narrow components at all (see Fig.~\ref{fig:compareoutburst} top).

\subsection{Gaussian fitting}\label{gaussian_fit}

\begin{figure}
\begin{center}
    \includegraphics[width=\columnwidth]{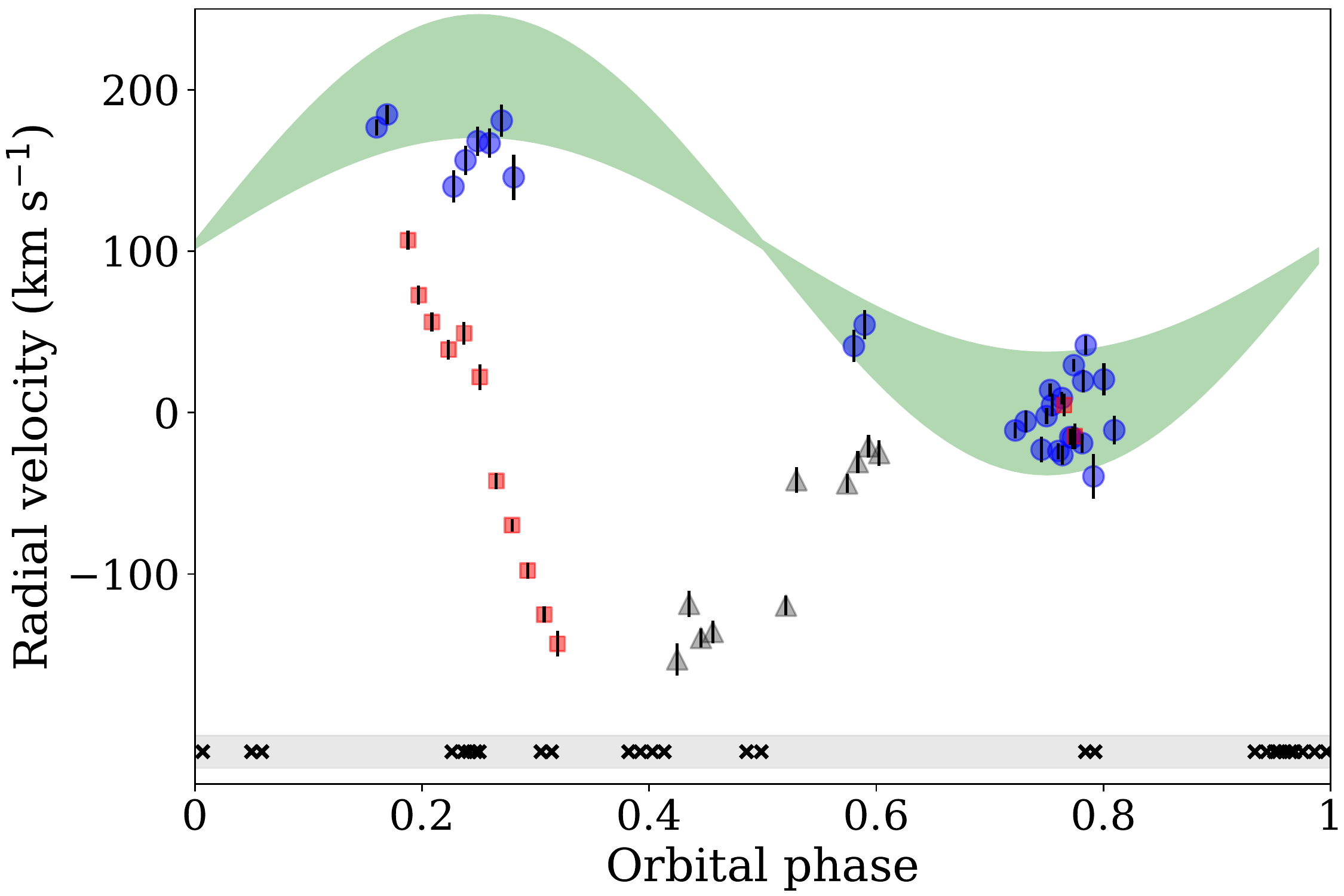}
\end{center} 
 \caption{Radial velocity as a function of orbital phase derived from Gaussian fits of the Bowen blend narrow components. The light green area represents the region where the radial velocity is consistent with that arising from the inner side of the companion star. Blue dots indicate velocities compatible (within 3$\sigma$) with the expected radial velocity of the donor, while gray triangles correspond to those that are not consistent. Crosses show the phases of those spectra in which N \textsc{iii} lines could not be identified. The velocities obtained from the VLT data \citep{2007MNRAS.375.1463C} are shown as red squares. }
    \label{fig:vel_all}
 
\end{figure}

Previous studies of several objects have shown that narrow components within the Bowen blend arise from the irradiated side of the companion star. Thus, $K_{\mathrm{em}}<K_{2}$ can in principle be obtained by directly measuring their radial velocity curve over a full orbital period \citep[e.g.][]{2002ApJ...568..273S}. To this end, we carried out Gaussian fits to the Bowen narrow components on each of the 38 spectra in which we clearly identify N \textsc{iii} narrow lines. All the spectra were initially normalised, to then fit the Bowen profile using a 3-Gaussian model. Two narrow components were used to fit the N \textsc{iii} 4634.13 \AA\ and  N \textsc{iii} 4640.64/4641.84 \AA\ emission lines. A broad  component was added in order to model the underlying broad Bowen blend profile. The position and height of each Gaussian were set as free parameters, but the relative separation between both narrow components was imposed to be constant in order to obtain a single and more stable velocity solution. The FWHM of the Gaussians were fixed to 200 and 1150 km s$^{-1}$ for the narrow and broad components, respectively. Thus, 5 parameters were optimized  through least-squares fitting, yielding a velocity offset and its associated error for every spectrum. In order to compute the orbital binary phases we used the  ephemeris from \citetalias{2017MNRAS.464L..41M}:  $T_{0}$ = 2455810.387$\pm$0.005 d and $P_{\mathrm{orb}}$  = 0.7895126 $\pm$0.000 0010 d. The velocity offset as a function of the orbital phase is shown in Fig. \ref{fig:vel_all}. We cannot clearly identify orbital motion of the companion; instead, a more complex distribution is observed (dots and triangles in Fig. \ref{fig:vel_all}).

Taken into account the system parameters given in \citetalias{2017MNRAS.464L..41M}, we used $K_{\mathrm{c}}$ for the case of low orbital inclination to compute limiting cases for the expected $K_{\mathrm{em}}$ in Aql~X-1. Even though a clear radial velocity curve was not observed, 28 points are consistent with the expected radial velocity, i.e. less than 3$\sigma$ from the region defined by the $K_{\mathrm{em}}$ limit cases.  The same analysis was applied to the Very Large Telescope (VLT) data presented in \citet{2007MNRAS.375.1463C}. The phase-dependent velocity offsets derived from these data are consistent with those originally reported (we consider now the \citetalias{2017MNRAS.464L..41M} ephemeris), but are not compatible with those expected for the companion star. Thus, our Gaussian fitting suggests that the Bowen blend is tracing  more than one emission region.
 
\subsection{ Doppler mapping}\label{dopplermap}

Doppler tomography \citep{1988MNRAS.235..269M} inverts phase-resolved data to obtain a brightness distribution in velocity space (i.e. Doppler coordinates). This technique is suitable to search for emission features from the companion star in faint systems or those for  which the spectra are contaminated by other components. The Doppler coordinate frame co-rotates with the system, with the origin placed on its centre of mass.  The $V_{\mathrm{x}}$-axis is defined by the line that joins the compact object with the companion star, while the $V_{\mathrm{y}}$-axis is set in the direction of the velocity of the companion. Thus, when the correct systemic velocity ($\gamma$) is used as the input to the Doppler code, the centre of mass of the system is placed at the origin on the Doppler map [i.e. ($V_{\mathrm{x}}$,$V_{\mathrm{y}}$)=(0,0)]. The emission from the donor should appear as a compact spot along the  $V_{\mathrm{y}}$ axis and  $K_{\mathrm{em}}$ can be determined from the position of the centroid of this spot if spectra are folded on the correct ephemeris. Doppler tomography makes use of spectra that ideally sample the orbital period in full. It uses all at once, being able to separate out the different emission sources. 

\begin{figure}
\begin{center}
    \includegraphics[width=\columnwidth]{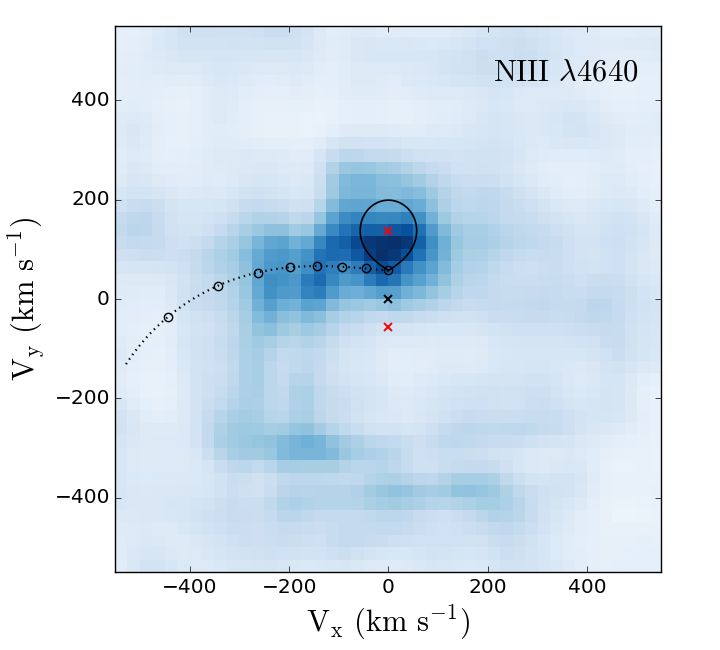}
\end{center}
 
 \caption{The Doppler tomogram of N \textsc{iii} 4640.64/4641.84  \AA\  computed using the updated binary period, ephemeris and $\gamma$ = 104 km $\mathrm{s^{-1}}$ from \citetalias{2017MNRAS.464L..41M}. The origin (denoted by a black cross) corresponds to the centre of mass of the system. We overplot the gas-stream trajectory and the Roche lobe of the companion assuming $K_{2}$ = 136 km $\mathrm{s^{-1}}$ and a mass ratio of 0.41. The true orbital velocities of the companion (0, $K_{2}$ = 136 km $\mathrm{s^{-1}}$) and the neutron star (0, -$\mathrm{K_1}$ = -56 km $\mathrm{s^{-1}}$) are denoted by red crosses. }
    \label{fig:doppler}

\end{figure}

Following \cite{2017MNRAS.466.2261W}, we used all our 65 spectra to compute Doppler maps. These were constructed by using the second generation (\textsc{python/c++} based), maximum entropy Doppler tomography code\footnote{This code is available at https://github.com/trmrsh/trm-doppler, see {\sc docs/features.rst} for its notable features.} developed by T. Marsh. We computed the Doppler tomogram for the strongest Bowen narrow component, N \textsc{iii} 4640.64/4641.84  \AA\, using the ephemeris from \citetalias{2017MNRAS.464L..41M} and also their systemic velocity  ($\gamma = 104 \pm 3$ km s$^{-1}$). We present the resulting Doppler map in Fig. \ref{fig:doppler}, where a compact spot along the positive $V_{\mathrm{y}}$ is clearly detected. We computed the position of this spot via 2D-Gaussian fitting and found $ V_{x} \sim -39$ km s$^{-1}$,  $V_{\mathrm{y}} \sim 94$ km s$^{-1}$. This corresponds to a phase shift of $\Delta \phi \sim -0.06 $, which is compatible with the irradiated donor star (see Fig. \ref{fig:doppler}; the angular size of the donor is $\sim 0.1$ in phase, while the absolute error in the orbital phase is only $\sim 0.01$). We also detected a bright extended region, which is consistent with the gas-stream trajectory and/or the region where the infalling material impacts the outer edge of the disc (i.e. the hot spot).

\begin{figure}
\begin{center}
   \includegraphics[width=\columnwidth]{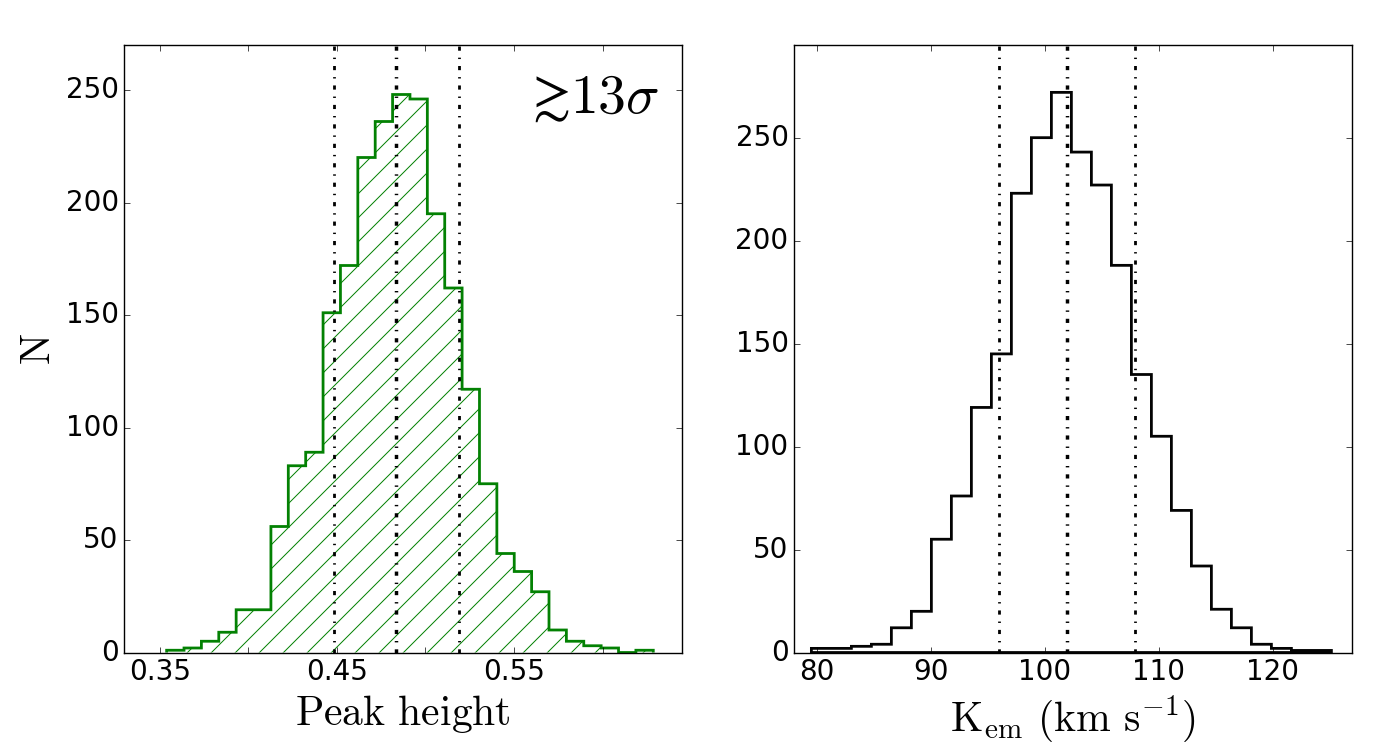}
\end{center}
 
 \caption{Distributions of the relative peak height (left) and $K_{\mathrm{em}}$ (right) determined from 2000 bootstrapped maps iterated to reach the same image entropy (S = -0.04) as that of the original image (Fig. \ref{fig:doppler}). Dashed lines indicate the mean and the $\pm$ 1$\sigma$ confidence intervals. The emission feature is significant at the $\gtrsim$13$\sigma$ level, providing a radial velocity semi-amplitude of $K_\mathrm{{em}}$ = 102 $\pm$ 6 km $\mathrm{s^{-1}}$.}
    \label{fig:dopplerdist}

\end{figure}

In a second step, we used bootstrap techniques to compute confidence intervals for the above results. To this end, we model a data set of 2000 bootstrapped maps from the original data  \citep[see][]{2017MNRAS.466.2261W}. We computed the peak height and $K_{\mathrm{em}}$ via a 2D Gaussian fit to each bootstraped map. Subsequently, histograms of both parameters were constructed. Since these bootstrap distributions were roughly Gaussian, we estimated the mean and the 1$\sigma$ error for each parameter via Gaussian fitting  (Fig. \ref{fig:dopplerdist}). We found $K_{\mathrm{em}}$ = 102 $\pm$ 6 km $\mathrm{s^{-1}}$. Additionally, we estimated the significance of the spot to be $\gtrsim$13$\sigma$.

\subsection{Opening angle of the accretion disc}\label{opangle}

The  $K_{\mathrm{em}}$ and  $K_{2}$ projected radial velocities can be related using $ K_{\mathrm{c}}$ ($=K_{\mathrm{em}}/K_{2}$) from \citetalias{2005ApJ...635..502M}. This correction mainly depends on the mass ratio ($q$) and the opening angle of the accretion disc ($\alpha$), which partially shadows the companion star, preventing the closest regions to the inner Lagrangian point from being irradiated. Given that $K_{\mathrm{em}}$, $K_{2}$ and $q$ are constrained, we carried out a Monte Carlo analysis to determine the range of accretion disc opening angles compatible with the observables. We proceeded as follows:

(i) We sampled $K_\mathrm{em}$ and $K_{2}$ by using Gaussian distributions from the values reported in Table \ref{parameters}. We created a synthetic distribution of $K_{\mathrm{c}}=K_{\mathrm{em}}/K_{2}$, that is,  we obtained a $K_{c}$ value for each combination of the two observables.

(ii) We produced synthetic $K_{\mathrm{c}}$ values from \citetalias{2005ApJ...635..502M}. These can be approximated by fourth-order polynomials on $q$  using the form:

\begin{table}
\centering
\caption{Observed values}
\label{parameters}
\begin{tabular}{cc} 
		\hline
		Parameter & Value \\
		\hline
		$ K_{\mathrm{em}}$ & 102 $\pm 6 $ km s$^{-1}$ \\
		$ K_{2}$ & 136 $\pm 4 $ km s$^{-1}$* \\		
		$ q$ & 0.41 $\pm 0.08 $ *\\	
		$ i$ & $36^{\circ}-47^{\circ}$*\\					
		\hline
		*\cite{2017MNRAS.464L..41M}		
		\label{parameters}
	\end{tabular}
\end{table}

\begin{equation}
K_{\mathrm{c}}\simeq N_{0}+N_{1}q+N_{2}q^{2}+N_{3}q^{3}+N_{4}q^{4}
\label{Kcorr}
\end{equation}

where $N_{0}...N_{4}$ are tabulated as a function of $\alpha$ for high and low inclination systems (90 deg and 40 deg, respectively). Based on \citetalias{2017MNRAS.464L..41M}, we considered the low inclination case. 

(iii) Using the value reported  in Table \ref{parameters}, we adopted a Gaussian distribution for $q$. This distribution was sampled from 0.05 to 0.85 in steps of 0.01; likewise, $\alpha$ was sampled from 0 to 22 deg in steps of 0.5 deg. For those $\alpha$ values not tabulated in  \citetalias{2005ApJ...635..502M}, we performed a second-order polynomial interpolation using the  $K_{\mathrm{c}}$ corresponding to the three nearest tabulated values, keeping the correction accurate to $\leq$ 2 per cent.
 
\begin{figure}
\begin{center}
    \includegraphics[width=\columnwidth]{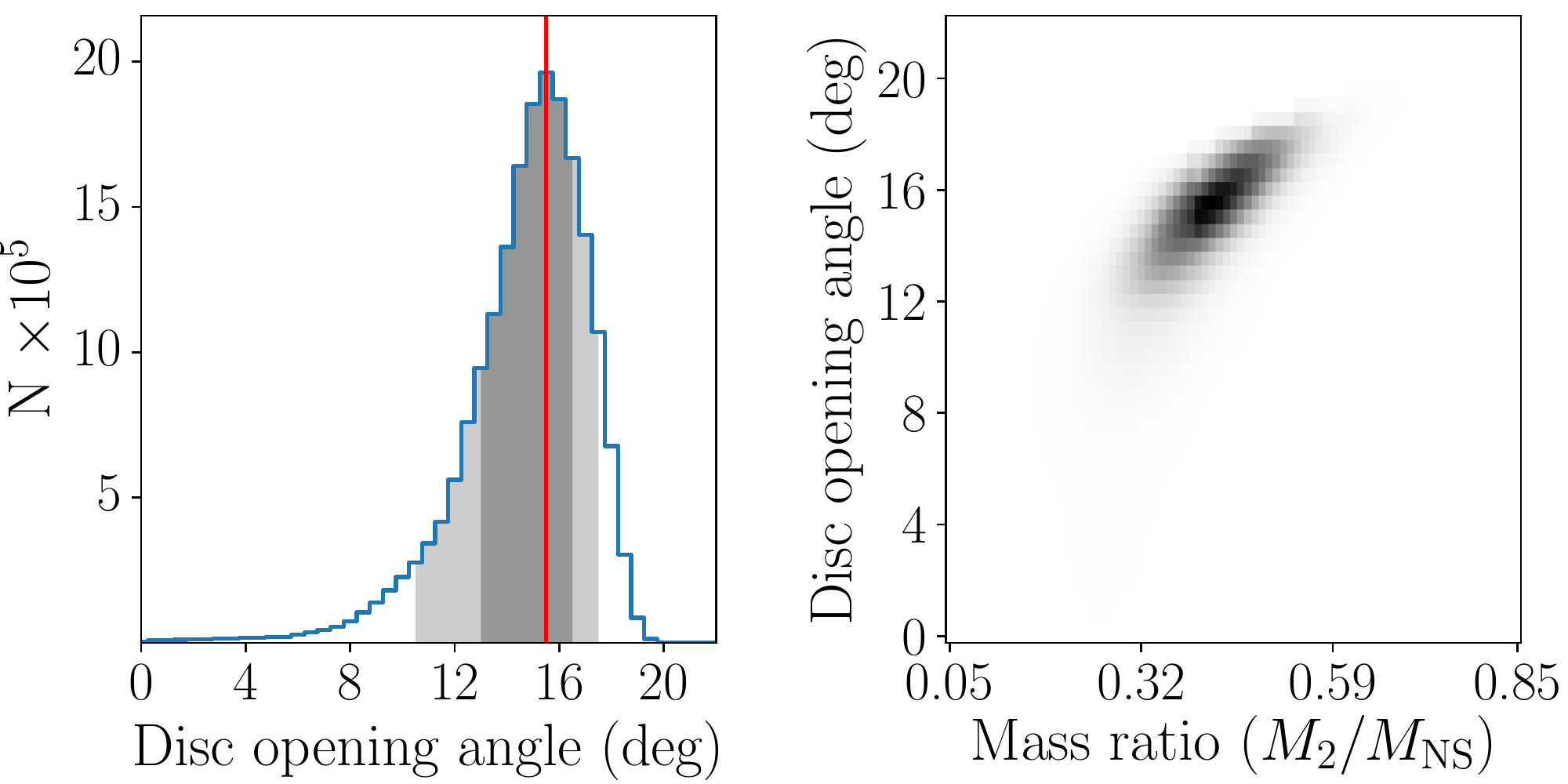}

\end{center}
 
 \caption{ Probability density functions obtained from the Monte Carlo analysis method with  $8\times10^{7}$ trials. The red line indicates the position of the 50th percentile. The 90 per cent and 68 per cent confidence regions are shown as dark and light shaded areas respectively. }
    \label{fig:alpha}

\end{figure}

(iv) We produced the probability density function of $\alpha$ by comparing each $K_{\mathrm{c}}$ Monte Carlo trial with the synthetic values for a given $q$ and within the $\alpha$ boundaries given by the Paczy\'nski approximation [i.e. $0<\alpha<\alpha_{M}$; where $\sin\, \alpha_{M}\cong 0.462 ( q/1+q)^{1/3}$ \cite{1971ARA&A...9..183P}]. The projected 1D probability density function of the disc opening angle was obtained using $8\times10^{7}$ Monte Carlo trials and then marginalized over $q$ (Fig. \ref{fig:alpha}). Considering 90 per cent confidence levels we obtained $\alpha = 15.5$ $ ^{+ 2.5}_{-5}$ deg.

\section{Discussion}\label{sec:discussion}

We have presented results from a detailed optical spectroscopic campaign of the neutron star transient Aql X-1 obtained during 3 different outbursts. The analysis is focussed on the Bowen blend spectral region and, in particular, we present a Doppler map of the N \textsc{iii} 4640.64/4641.84  \AA\ emission line, which shows a highly significant spot at the expected Doppler coordinates of the donor star. The $K_{\mathrm{em}}$ derived from this map is fully compatible with the true $K_{2}$ velocity reported in \citetalias{2017MNRAS.464L..41M}. The Gaussian-fitting technique, on the other hand, yields very complex results, which are in many cases inconsistent with a companion star origin. We interpret this as being the result of additional components that we are not able to fully resolve and appear blended with the emission arising from the irradiated side of the donor. In addition, the largely differing Bowen blend profiles observed over the observing campaign (Fig. \ref{fig:compareoutburst} top) shows that the relative contribution of the different emission regions contributing to the Bowen blend change with time. 

The $K_{\mathrm{em}}$ proposed in \cite{2007MNRAS.375.1463C} using the Bowen technique is ruled out by the $K_{2}$ velocity reported in \citetalias{2017MNRAS.464L..41M}. The spectral resolution of the former work is $\sim$ 100 km $\mathrm{s^{-1}}$ (i.e. significantly better than 130 --170 km $\mathrm{s^{-1}}$ used in the present work), but the data covered only a limited orbital phase range ($\sim$0--0.3) with all the spectra taken over 3 consecutive nights. Indeed, our re-analysis of the VLT data (which includes the new ephemeris; Fig \ref{fig:vel_all}) shows that none of the velocity measurement from this dataset were compatible with a donor star origin. Therefore, at least for the particular epoch and orbital phases covered by the VLT 2004 observations, the Bowen blend narrow emission was not dominated by the companion star contribution. These results together with the recently reported study of the black hole transient GX~339-4 \citep{2017arXiv170804667H} demonstrate that complete phase coverage is a mandatory requirement in order to be able to confidently apply the Bowen technique. A complete phase coverage also allows the application of more robust analysis techniques, such as Doppler tomography, which enables us to be able to disentangle the different emission regions and single out the companion star contribution. In this particular case, different emission regions are resolved in velocity space, including a very significant companion star contribution ($\gtrsim$13$\sigma$). Misidentification effects were minimized due to the amount and time distribution of the data, as we used 65 spectra taken over 3 different outbursts at different orbital phases. We obtained $K_{\mathrm{em}}$ = 102 $\pm$ 6 km $\mathrm{s^{-1}}$, which implies  $K_{\mathrm{c}} \sim$ 0.75.  This result is in the expected range, as computed by \citetalias{2005ApJ...635..502M}. 
Using the constraints to the orbital inclination and the mass ratio provided in \citetalias{2017MNRAS.464L..41M}, we ran Monte Carlo simulations to determine the opening angle of the accretion disc, $\alpha = 15.5$ $ ^{+ 2.5}_{-5}$ deg, where errors are given at 90 per cent significance level. We note that this value is, in principle, independent of any assumed disc geometry, as it just represents the angle subtended by the largest vertical structure of the accretion disc. It can be easily interpreted as the angle subtended by the outer disc rim but also as that related to a more complex structure \citep[e.g. a inner disk torus,][]{2013Sci...339.1048C}.

Several works have attempted to constrain the accretion disc opening angle by modelling optical fluxes and light curves (from photometric data). Considering a simple geometric model for X-ray reprocessing on LMXBs and comparing the absolute visual magnitude with the amplitude of the outburst light curve, \cite{1996A&A...314..484D} proposed an average value of $\alpha \sim$ 12 deg. Likewise, \cite{1976ApJ...209..562G} obtained $\alpha \sim 10$ deg for Hercules X-1 whilst \cite{1987ApJ...313..792M} estimated $\alpha\sim$ 9--13 deg for XB 1254-690. On purely theoretical grounds, \cite{1982A&A...106...34M} studied the role of X-ray heating on the vertical structure of the accretion disc and showed that X-ray radiation is expected to thicken the accretion disc significantly. They obtained $\alpha\sim$6 deg using models with no X-ray heating, while a thicker disc with $\alpha\sim$18--22 deg was determined when X-ray irradiation was taken in account. However, the latter value is probably overestimated as the disc is assumed to absorb every X-ray photon (see also \citealt{1996A&A...314..484D}).

In this paper we have presented the first solid determination of the accretion disc opening angle through a (relatively direct) spectroscopic method. It directly compares day-side and centre of mass velocities to infer the size of the accretion structure that shadows the donor. Even though we have carried out an accurate study (including detailed error analysis), the reader must bear in mind that the value presented here depends on the K-correction modelling (i.e. \citetalias{2005ApJ...635..502M}) and therefore is likely subject to systematic effects. Nevertheless, our results are clearly consistent with an irradiation-driven thick accretion disc, providing strong support for the aforementioned theoretical works. Values lower than $\sim 10$ deg seem very difficult to reconcile with our results, while the upper limit is roughly consistent with the maximum disc opening angle for the donor to be irradiated at all (whose value depends on $q$); see section \ref{opangle}. Indeed, opening angles close to this limit might explain the complex and very variable evolution of the Aql X-1 Bowen blend narrow components. On one hand, emission from the donor would be weak in this case and, on the other hand, variable X-ray irradiation could produce small changes in the vertical structure of the disc, which might, in some cases, go beyond the previous limit, preventing the donor from being irradiated. In light of this, we searched for trends relating the strength/presence of narrow emission lines with the X-ray flux measured by the \textit{MAXI} monitor (Fig. \ref{fig:compareoutburst}) and found no conclusive evidence. 

Finally, we remark that our results are valid for the outburst state, as thinner accretion discs are expected in quiescence (see above). To test this scenario, we carried out the Monte Carlo analysis presented here using literature data from the neutron star transient Centaurus X-4. We considered  $K_{\mathrm{em}}$= 122.8$\pm$ 11.8 km $\mathrm{s^{-1}}$ \citep[obtained from a Doppler map of the HeI 5876~\AA\ emission line while the source was in quiescence;][]{2005A&A...444..905D},  $K_{2}$= 144.6$\pm$ 0.3 km $\mathrm{s^{-1}}$ and  $q$=0.20 $\pm$ 0.03  \citep[from][]{2007A&A...470.1033C}. We obtained $\alpha = 12.5$ $ ^{+ 2.5}_{-6.5}$ deg  (errors are given at 90 per cent), which is consistent with 7 deg $\leq\alpha\leq$14 deg  determined by \cite{2006A&A...460..257D} by directly using the $1\sigma$ limits to the radial velocity curves and mass ratio (we note that our 68 per cent constraint is $\alpha = 12.5$ $ ^{+1.5}_{-3}$) deg. Therefore, even if this measurement allows for smaller accretion disc opening angles, it is not significantly different from the outburst value presented here.

\section{Conclusions}\label{sec:conclusions}

We have presented time-resolved spectroscopy of Aql X-1 in outburst with an unprecedented orbital phase coverage. We resolved Bowen blend narrow components and proved that the Bowen technique is able to trace the orbital motion of the companion star when ample phase coverage is available. We used new generation Doppler mapping to determine $K_{\mathrm{em}}$ from N \textsc{iii} 4640.64/4641.84  \AA\ emission line. This, together with $K_{2}$ and $q$ values from near-infrared spectroscopy, allowed us to directly measure the elusive accretion disc opening angle in outburst ($\alpha = 15.5$ $ ^{+ 2.5}_{-5}$ deg), the large value of which is in agreement with accretion disc models that take into account X-ray irradiation.

\section*{Acknowledgements}

We are thankful to the anonymous referee for providing useful comments that improved the manuscript. Based on data from the GTC Public Archive at CAB (INTA-CSIC), developed in the framework of the Spanish Virtual Observatory project supported by the Spanish MINECO through grants AYA 2011-24052 and AYA2014-55216. The system is maintained by the Data Archive Unit of the CAB (INTA-CSIC). DMS acknowledges Fundaci\'{o}n La Caixa for the financial support received in the form of a PhD contract. TMD is supported by RYC-2015-18148. \textsc{molly} software developed by T. R. Marsh is gratefully acknowledged.




\bibliographystyle{mnras}
\bibliography{biblio}




\bsp	
\label{lastpage}
\end{document}